\documentstyle[11pt]{article}
\normalbaselines
\newcounter{eqnletter}[equation]
\catcode`\@=11
\@addtoreset{equation}{section}   
\catcode`\@=12

\begin{document}

\begin{center}

{\LARGE\bf Quasi-Exactly Solvable Deformations of Gaudin
Models and ``Quasi-Gaudin Algebras''.}\\[0.7cm]

\vskip 1cm

{\large {\bf A.G. Ushveridze} }

\vskip 0.1 cm

Department of Theoretical Physics, University of Lodz,\\
Pomorska 149/153, 90-236 Lodz, Poland\footnote{E-mail
address: alexush@krysia.uni.lodz.pl}

\end{center}
\vspace{1 cm}
\begin{abstract}

A new class of completely integrable models is constructed.
These models are deformations of the famous integrable and
exactly solvable Gaudin models. In contrast with the latter, 
they are quasi-exactly solvable, i.e. admit the algebraic
Bethe ansatz solution only for some limited
parts of the spectrum. An underlying algebra responsible
for both the phenomena of complete integrability and 
quasi-exact solvability is constructed. We call it
"quasi-Gaudin algebra" and demonstrate that it is a special 
non-Lie-algebraic deformation of the ordinary Gaudin algebra.

\end{abstract}

\newpage 

\section{Introduction}
\label{0}

A quantum model is usually called {\it exactly solvable} if all
solutions of its spectral problem can be found algebraically.
Recently a new important class of the so-called
{\it quasi-exactly solvable} models has been discovered 
\cite{zas,turush,ush,tur}.
These models are distinguished by the fact that only 
certain limited parts of their spectra admit an algebraic construction.
A detailed exposition of different theories explaining the
phenomenon of quasi-exact solvability and proposing
constructive methods for building and solving such models can be found in 
papers \cite{ushrev1,shifrev1,moretal,GKO1,GKO2,ushrev2, shifrev2}
and in the book \cite{ushbook}.

The quasi-exactly solvable models can be considered as {\it
deformations} of exactly solvable ones \cite{ushbook}. 
Usually, the deformation of a
given exactly solvable model leads to an infinite sequence of
quasi-exactly solvable models with different hamiltonians
and different number of exact solutions \cite{ushbook}.

For example, the simplest exactly solvable model of a multi-dimensional 
harmonic oscillator with hamiltonian
\begin{eqnarray}
H=-\sum_{i=1}^d\frac{\partial^2}{\partial
x_i^2}+\sum_{i=1}^d b_i^2 x_i^2
\label{0.1}
\end{eqnarray}
can be deformed into an infinite sequence of quasi-exactly solvable 
models of
multi-dimensional sextic anharmonic oscillators with hamiltonians
\begin{eqnarray}
H_n=-\sum_{i=1}^d\frac{\partial^2}{\partial
x_i^2}+\sum_{i=1}^d b_i^2 x_i^2 +2ar^2\left[\sum_{i=1}^d b_i x_i^2
-2n-1-\frac{d}{2}\right]+a^2 r^6, \quad n=0,1,\ldots.
\label{0.2}
\end{eqnarray}
In contrast with (\ref{0.1}), each of models (\ref{0.2}) 
has only $(n+d)!/(n!d!)$ algebraically constructable solutions \cite{ush}.
Here the role of a deformation parameter is played by $a$.

It is worth stressing that both models (\ref{0.1}) and
(\ref{0.2}) are completely integrable in the sense that
they have enough number of mutually commuting integrals of
motion (see e.g. \cite{ushbook}).
Thus, the above example clearly demonstrates that the deformation 
procedure may preserve the integrability property. 
There are many other examples of such a sort which can be
found in ref. \cite{ushbook}.

It is natural to assume that any completely integrable and exactly
solvable quantum model can be deformed into an infinite set of integrable 
and quasi-exactly solvable models. If this assumption is true, then the 
following interesting problem immediately arises.

\medskip
{\bf The Problem.} It is well known that most of physically interesting
{\it integrable} and {\it exactly solvable} models can be obtained and
solved in the framework of the celebrated $R$-matrix approach 
\cite{kor,fad1,fad2}. A mathematical structure lying in the ground of 
this approach and  responsible for both the phenomena of integrability 
and exact solvability is the so-called Yang -- Baxter (YB) algebra 
(see e.g. the book \cite{jimbo}). Using
its generators, one can easily construct not only all the
integrals of motion of a model (and thus ensure its integrability)
but also build all eigenvectors and eigenvalues of the corresponding
spectral problem (and thus ensure its exact solvability in
the framework of the so-called Algebraic Bethe Ansatz
\cite{fad2}).
Let us now consider any quasi-exactly solvable deformation of such a model
that preserves its integrability property. 
If such a deformation does exist, then it is natural to ask 
ourselves if there is any algebra responsible 
for both the phenomena of {\it integrability} and 
{\it quasi-exact solvability} of
the deformed model? In other words, is there any analogue
of the YB algebra (some kind of "quasi-YB algebra")
whose generators could be used not only for constructing
the integrals of motion of the arising quasi-exactly solvable models
but also for building their algebraically constructable
eigenvalues and eigenvectors? We mean here some analogue of
the Algebraic Bethe Ansatz. And the last and, may be, the most
important question: is it possible to interpret this
quasi-YB algebra as a {\it deformation} of the ordinary YB algebra?  

\medskip
It is obvious that if solution of this problem
does actually exist and the quasi-YB algebras will be found, then, 
developing a
regular method for building their representations we obtain 
a big factory of new integrable and quasi-exactly solvable models. Since
the ordinary YB algebras give rise to many physically 
interesting exactly solvable
systems of quantum mechanics, statistical physics and field theory, 
it would be natural to expect that the set of quasi-exactly solvable 
models associated with quasi-YB algebras also will contain a lot of 
physically
interesting systems.

\medskip
It seems that the posed problem has a positive solution. In this
paper we consider the simplest case of the so-called 
$sl(2)$-Gaudin algebra $G[sl(2)]$ which is closely related to the 
classical YB
algebra and leads to the completely integrable Gaudin models
solvable by means of Bethe ansatz \cite{gau,jur} (section \ref{1}). 
In sections \ref{2} and \ref{3} we construct a special
non-Lie algebraic deformation of this algebra which automatically produces
an infinite series of integrable models. We call this
algebra the "quasi-$G[sl(2)]$ algebra" and the corresponding models
-- the "quasi-Gaudin models". In sections \ref{4}, \ref{5}
and \ref{6} we show that all these models are quasi-exactly solvable in 
the sense that they admit only a partial Bethe ansatz solution of 
spectral problem. In section \ref{7} we
construct a simple realization of quasi-$G[sl(2)]$ algebra and
later, in section \ref{8}, demonstrate in an independent way 
that quasi-Gaudin models associated with this concrete
realization are actually quasi-exactly solvable.
In the last section \ref{9} we introduce a concept of quasi-$sl(2)$
algebra which can be viewed as a special limiting case of
quasi-$G[sl(2)]$.

\section{The Gaudin algebra}
\label{1}

In this section we remind the reader some basic facts concerning 
Gaudin algebras, their representations and properties of the associated
Gaudin models. More detailed exposition of this subject can
be found in refs. \cite{gau,jur,ushrev2,ushbook}.

\medskip
The Gaudin algebra $G[sl(2)]$ is an infinite-dimensional
extension of the ordinary $sl(2)$ algebra.
Its three generators, 
$\vec S(\lambda)=\{S^-(\lambda), S^0(\lambda),
S^+(\lambda)\}$ with
$\lambda\in C$ obey the commutation relations
\begin{eqnarray}
S^0(\lambda)S^0(\mu)
-S^0(\mu)S^0(\lambda)&=&0, \nonumber\\[0.4cm]
S^\pm(\lambda)S^\pm(\mu)
-S^\pm(\mu)S^\pm(\lambda)&=&0, \nonumber\\[0.3cm]
S^0(\lambda)S^\pm(\mu)
-S^\pm(\mu)S^0(\lambda)&=&
\pm\frac{S^\pm(\lambda)-S^\pm(\mu)}{\mu-\lambda},\nonumber\\
S^-(\lambda)S^+(\mu)
-S^+(\mu)S^-(\lambda)&=&
2\frac{S^0(\lambda)-S^0(\mu)}{\mu-\lambda},
\label{1.1}
\end{eqnarray}
generalizing those of $sl(2)$. Using 
(\ref{1.1}), one can easily prove that the operators
\begin{eqnarray}
C(\lambda)=S^0(\lambda)S^0(\lambda)-\frac{1}{2}
S^-(\lambda)S^+(\lambda)-\frac{1}{2}S^+(\lambda)S^-(\lambda)
\label{1.2}
\end{eqnarray}
form a commutative family,
\begin{eqnarray}
[C(\lambda),C(\mu)]=0,
\label{1.3}
\end{eqnarray}
and thus, can be interpreted as integrals
of motion of a certain quantum completely 
integrable model. The latter is known under name of Gaudin model.

The role of the ``Hilbert space'' in which the
operators $C(\lambda)$ act is played by the 
representation space of Gaudin algebra.
In order to construct it one needs to fix the lowest weight vector
$|0\rangle$ and the lowest weight function $f(\lambda)$
obeying the relations
\begin{eqnarray}
S^0(\lambda)|0\rangle=f(\lambda)|0\rangle, \qquad
S^-(\lambda)|0\rangle=0.
\label{1.4}
\end{eqnarray}
After this, we can define the representation space as
a linear hull of vectors
\begin{eqnarray}
|\xi_1,\ldots,\xi_m\rangle=S^+(\xi_m)\ldots S^+(\xi_1)|0\rangle
\label{1.5}
\end{eqnarray}
with arbitrary $m$ and $\xi_1,\ldots,\xi_m$. We shall
denote this representation space by $W_{f(\lambda)}$.

The ``Schr\"odinger equation'' for the Gaudin model reads now
\begin{eqnarray}
C(\lambda)\phi=E(\lambda)\phi, \quad \phi\in W_{f(\lambda)}.
\label{1.6}
\end{eqnarray}
The beauty of this equation lies in the fact that all
its solutions can be obtained algebraically.
For example, using commutation relations (\ref{1.1}) and
formulas (\ref{1.4}) it is easy to check that the lowest
weight vector is always a solution of the Gaudin problem:
\begin{eqnarray}
C(\lambda)|0\rangle=\left(f^2(\lambda)+\frac{\partial}{\partial \lambda}
f(\lambda)\right)|0\rangle.
\label{1.7}
\end{eqnarray}
The remaining solutions of equation (\ref{1.6}) 
can be obtained by means of the so-called algebraic Bethe ansatz
\begin{eqnarray}
\phi=S^+(\xi_m)S^+(\xi_{m-1})\ldots S^+(\xi_2)S^+(\xi_1)|0\rangle,
\label{1.8}
\end{eqnarray}
in which $m$ is an arbitrary non-negative integer and
$\xi_1,\ldots,\xi_m$ are some unknown numbers. It can be
demonstrated that vector (\ref{1.8}) is an eigenvector of
the Gaudin operator with the eigenvalue 
\begin{eqnarray}
E(\lambda)=f^2(\lambda)+\frac{\partial}{\partial\lambda}f(\lambda)
+\sum_{i=1}^m 
\frac{f(\lambda)-f(\xi_i)}{\lambda-\xi_i}
\label{1.9}
\end{eqnarray}
if the following conditions hold
\begin{eqnarray}
\sum_{k=1,k\neq i}^m \frac{1}{\xi_i-\xi_k}+f(\xi_i)=0,\quad
i=1,\ldots, m.
\label{1.10}
\end{eqnarray}
These conditions are known as Bethe ansatz equations.
It is known that the constructed solutions with $m=0,1,\ldots$ 
expire all possible solutions of the Gaudin spectral
problem and therefore the latter is {\it exactly solvable}.

\section{The quasi-Gaudin algebra}
\label{2}

In this section we consider a special modification of Gaudin
algebra which also leads to completely integrable quantum systems.

\medskip
Let $\vec S_n(\lambda)=\{S_n^-(\lambda), S_n^0(\lambda), S_n^+(\lambda)\}$,
$n\in Z$, $\lambda\in C$ denote the operators obeying the relations
\begin{eqnarray}
S_n^0(\lambda)S_n^0(\mu)
-S_n^0(\mu)S_n^0(\lambda)&=&0, \nonumber\\[0.4cm]
S_{n\pm 1}^\pm(\lambda)S_n^\pm(\mu)
-S_{n\pm 1}^\pm(\mu)S_n^\pm(\lambda)&=&0, \nonumber\\[0.3cm]
S_{n\pm 1}^0(\lambda)S_n^\pm(\mu)
-S_n^\pm(\mu)S_n^0(\lambda)&=&
\pm\frac{S_n^\pm(\lambda)-S_n^\pm(\mu)}{\mu-\lambda},\nonumber\\
S_{n+1}^-(\lambda)S_n^+(\mu)
-S_{n-1}^+(\mu)S_n^-(\lambda)&=&
2\frac{S_n^0(\lambda)-S_n^0(\mu)}{\mu-\lambda}.
\label{2.1}
\end{eqnarray}
We consider (\ref{2.1}) as the defining relations of
a certain infinite-dimensional algebra. It
is not difficult to see that this algebra is very similar to
the Gaudin algebra but, in contrast with the latter, it {\it is
not} a Lie algebra\footnote{In section \ref{7} we demonstrate that 
(\ref{2.1}) can be viewed as a deformation of the Gaudin algebra.}.
We call it the ``quasi-Gaudin algebra'' or, more concretely,
``quasi-$G[sl(2)]$''.

The algebra (\ref{2.1}) has many remarkable properties.
First of all, only from the ``quasi-commutation relations'' (\ref{2.1})
it immediately follows that the operator valued functions
\begin{eqnarray}
C_n(\lambda)=S_n^0(\lambda)S_n^0(\lambda)-\frac{1}{2}
S_{n+1}^-(\lambda)S_n^+(\lambda)-
\frac{1}{2}S_{n-1}^+(\lambda)S_n^-(\lambda)
\label{2.2}
\end{eqnarray}
form commutative families for any $n$:
\begin{eqnarray}
[C_n(\lambda),C_n(\mu)]=0.
\label{2.3}
\end{eqnarray}
These functions are obvious generalizations of the
Gaudin integrals of motion. Note, however, that the
commutators between $C_n(\lambda)$ and $C_m(\lambda)$ do
not vanish if $m\neq n$.

The property (\ref{2.3}) suggests to interpret
$C_n(\lambda)$ as generating functions of commuting
integrals of motion for certain completely integrable
quantum models. In fact, (\ref{2.3}) describes an
infinite sequence of such models, since the integer $n$ 
can be chosen arbitrarily.
This is the first important difference with the Gaudin case
when an analogous construction leads to a single completely
integrable model --- the Gaudin model.

Exactly as in the case of Gaudin algebra, the
construction of these models needs the specification of the
``Hilbert space''. It would be natural to identify it
with the representation space of our algebra.
The latter can be constructed in the same way as in the
Gaudin case. First of all, we need the lowest weight vector
$|0\rangle$ which, as in the Gaudin case, should be 
an eigenvector of both the lowering and
neutral operators $S_n^-(\lambda)$ and $S_n^0(\lambda)$.
The most general relations expressing this fact and
compatible with the quasi-commutation relations (\ref{2.1})
read 
\begin{eqnarray}
S_n^0(\lambda)|0\rangle=\left(F(\lambda)
+nG(\lambda)\right)|0\rangle,\qquad
S_n^-(\lambda)|0\rangle=nG(\lambda)|0\rangle,
\label{2.4}
\end{eqnarray}
where $F(\lambda)$ and $G(\lambda)$ are certain 
arbitrarily fixed functions\footnote{Stictly speaking, the
$G(\lambda)$ functions in the first and second formulas of
(\ref{2.4}) may differ from each other by a certain $n$-dependent
constant factor. However, using the transformations 
$S_n^0(\lambda)\rightarrow S_n^0(\lambda)$,
$S_n^-(\lambda)\rightarrow c_n S_n^-(\lambda)$,
$S_n^+(\lambda)\rightarrow c_{n+1}^{-1} S_n^+(\lambda)$ which
do not change the commutation relations
(\ref{2.1}), we can simply remove this difference.}.
Note that the lowest
weight vector $|0\rangle$ is not annihilated by the 
``lowering operators'' $S_n^-(\lambda)$ except for the case
of $n=0$. This fact, which will play a determining
role in our further considerations, gives us the second important
difference with the Gaudin case.

After fixing the ``lowest weight functions'' $F(\lambda)$ and
$G(\lambda)$, we can define the representation space 
as a linear hull of vectors 
\begin{eqnarray}
|\xi_1,\ldots,\xi_m\rangle_k=S_{m+k}^+(\xi_m)S_{m-1+k}^+(\xi_{m-1})
\ldots S_{2+k}^+(\xi_2) S_{1+k}^+(\xi_1)|0\rangle
\label{2.5}
\end{eqnarray}
with arbitrary $k$, $m$ and $\xi_1,\ldots,\xi_m$.
The convenience of this definition comes from the fact that, due to the
relations (\ref{2.1}), the vectors (\ref{2.5}) are symmetric with 
respect to
all permutations of numbers $\xi_1,\ldots,\xi_m$ (exactly
as in the Gaudin case).
We denote the representation space of algebra (\ref{2.1})
by $W_{F(\lambda),G(\lambda)}$. 

\section{The quasi-Gaudin model}
\label{3}

Now we are ready to construct the integrable models
associated with algebra (\ref{2.1}). We postulate that 
Schr\"odinger equations for these models read
\begin{eqnarray}
C_n(\lambda)\phi_n=E_n(\lambda)\phi_n, \quad \phi_n\in
W_{F(\lambda),G(\lambda)}, \qquad n=0,1,\ldots.
\label{3.1}
\end{eqnarray}
So, formula (\ref{3.1}) defines an infinite sequence of
integrable models, each of which is completely 
determined by a triple $\{F(\lambda), G(\lambda), n\}$.

\section{A trivial solution}
\label{4}

It is known that in the Gaudin case the lowest weight
vector is always a solution of the Gaudin spectral problem.
Is this true in the case of models (\ref{3.1})?
In order to answer this question, we should simply act by the
operator $C_n(\lambda)$ on the vector $|0\rangle$ and look
at the result. For this it is convenient to rewrite the
operator $C_n(\lambda)$ in a little bit different form.
We use the formula
\begin{eqnarray}
S_{n+1}^-(\lambda)S_n^+(\lambda)
-S_{n-1}^+(\lambda)S_n^-(\lambda) =
-2\frac{\partial}{\partial \lambda} S_n^0(\lambda),
\label{4.1}
\end{eqnarray}
which follows from the quasi-commutation relations
(\ref{2.1}) in the limit $\mu\rightarrow \lambda$ and, substituting it
into the expression (\ref{2.2}) for $C_n(\lambda)$, obtain
\begin{eqnarray}
C_n(\lambda)=S_n^0(\lambda)S_n^0(\lambda)+
\frac{\partial}{\partial\lambda} S_n^0(\lambda)
-S_{n-1}^+(\lambda)S_n^-(\lambda).
\label{4.2}
\end{eqnarray}
From (\ref{4.2}) and formulas (\ref{2.4}) it immediately
follows that
\begin{eqnarray}
C_n(\lambda)|0\rangle=
\left\{\left(F(\lambda)+nG(\lambda)\right)^2
+\frac{\partial}{\partial \lambda}
\left(F(\lambda)+nG(\lambda)\right)\right\}|0\rangle
+nG(\lambda)S_{n-1}^+(\lambda)|0\rangle.
\label{4.3}
\end{eqnarray}
Now it becomes clear that vector $|0\rangle$
could be a solution of problem (\ref{3.1})
only if $n=0$. In this case we obtain the relation
\begin{eqnarray}
C_0(\lambda)|0\rangle=
\left(F^2(\lambda)+\frac{\partial}{\partial\lambda}
F(\lambda)\right)|0\rangle
\label{4.4}
\end{eqnarray}
which is similar to the Gaudin relation (\ref{1.7}).

\section{The Bethe ansatz}
\label{5}

In analogy with the Gaudin case, let us try to find the
solutions of spectral equations (\ref{3.1}) by means of the
algebraic Bethe ansatz.

It is natural to try to take the Bethe vector in the form
\begin{eqnarray}
\phi_n=S_{m+k}^+(\xi_m)S_{m-1+k}^+(\xi_{m-1})
\ldots S_{2+k}^+(\xi_2) S_{k+1}^+(\xi_1)|0\rangle
\label{5.1}
\end{eqnarray}
with some $k$ and $m$, and, using the relations
\begin{eqnarray}
C_n(\lambda)S_{n-1}^+(\mu)
-S_{n-1}^+(\mu)C_{n-1}(\lambda) =
2\frac{S_{n-1}^+(\mu)S_{n-1}^0(\lambda)
-S_{n-1}^+(\lambda)S_{n-1}^0(\mu)}{\lambda-\mu},
\label{5.2}
\end{eqnarray}
which follow from the basic relations (\ref{2.1}),
try to transfer the operator $C_n(\lambda)$ to the right.
From formula (\ref{5.2}) it is seen that, in order to start
performing such a permutation, it is necessary to take 
\begin{eqnarray}
m+k=n-1. 
\label{5.3}
\end{eqnarray}
Now note that each permutation of the operator $C_n(\lambda)$ 
with raising generators decreases its index by one.
This means that after $m$ permutations when this operator 
appears in front of the
lowest weight vector, it will have the index $n-m$. 
The standard presciptions to Bethe ansatz technique imply
that the hamiltonian of an integrable system, appearing
after all permutations in front of the lowest weight vector, 
should be absorbed by it. But such an absorbtion is
possible only if the lowest weight vector is an eigenvector of 
a hamiltonian. According to the result of previous section, the lowest 
weight vector is an eigenvector of the operator
$C_{n-m}(\lambda)$ only if
\begin{eqnarray}
n-m=0. 
\label{5.4}
\end{eqnarray} 
Comparing formulas (\ref{5.3}) and {\ref{5.4}} we can conclude
that the only case when the
ansatz (\ref{5.1}) for equation (\ref{3.1}) may lead to
some algebraic solutions corresponds to the choice
\begin{eqnarray}
k=-1,\quad m=n.  
\label{5.5}
\end{eqnarray} 

\section{The Bethe ansatz solution}
\label{6}

After using the restrictions (\ref{5.5}), 
the ansatz (\ref{5.1}) takes the form
\begin{eqnarray}
\phi_n=S_{n-1}^+(\xi_n)S_{n-2}^+(\xi_{n-1})
\ldots S_1^+(\xi_2) S_0^+(\xi_1)|0\rangle.
\label{6.1}
\end{eqnarray}
Let us now check that it actually contains solutions of the problem
(\ref{3.1}). This can be demonstrated exactly in the same way as in the
ordinary Gaudin case.

Repeating the reasonongs of ref. \cite{gau}, we transfer the
operator $C_n(\lambda)$ to the right by using the
quasi-commutation relations (\ref{5.2}).
The neutral generators of quasi-Gaudin algebra appearing after such a 
transference also should be transfered to the right. 
Exactly as in the case of a $C$-operator, any permutation of a 
$S^0$-operator 
with raising generators forming the Bethe vector decreases
its index by one. Finally, after completing all possible transferences,
we obtain a lot of terms with operators 
$C_0(\lambda)$, $S_0^0(\lambda)$ and $S_0^0(\xi_i)$ 
standing in front of the lowest weight vector $|0\rangle$. After this one
should get rid of these operators by using formulas (\ref{2.4})
and (\ref{4.4}). The result has the form
\begin{eqnarray}
C_n(\lambda)S_{n-1}^+(\xi_n)\ldots S_0^+(\xi_1)|0\rangle=
A(\lambda,\xi_1,\ldots,\xi_n)S_{n-1}^+(\xi_n)\ldots S_0^+(\xi_1)|0\rangle+
\nonumber\\
\sum_{i=1}^n \frac{B_i(\xi_1,\ldots,\xi_n)}{\lambda-\xi_i}
S_{n-1}^+(\xi_n)\ldots S_{i}^+(\xi_{i+1})S_{i-1}^+(\lambda)
S_{i-2}^+(\xi_{i-1})\ldots S_0^+(\xi_1)|0\rangle,
\label{6.2}
\end{eqnarray}
where $A(\lambda,\xi_1,\ldots,\xi_n)$ and
$B_i(\xi_1,\ldots,\xi_n), \ i=1,\ldots,n$ are
some explicitly constructable functions.
We see that the right hand side of (\ref{6.2}) consists of
two parts. The first part is proportional to the Bethe
vector $\phi_n$, while the second one consists of $n$ terms
which do not have such a form (these are the so-called
"unvanted terms"). This means that the
coefficient function $A(\lambda,\xi_1,\ldots,\xi_n)$ 
determine the eigenvalues of the operator $C_n(\lambda)$
provided that the unwanted terms are absent. But in order
to get rid of the unwanted terms it is sufficient to
require the vanishing of their coefficients 
$B_i(\xi_1,\ldots,\xi_n)=0,\quad i=1,\ldots, n$
This leads us to a system of $n$ equations for $n$ unknowns 
$\xi_1,\ldots,\xi_n$, which is nothing else than a typical
system of Bethe ansatz equations. 
The explicit form of these equations reads
\begin{eqnarray}
\sum_{k=1,k\neq i}^n \frac{1}{\xi_i-\xi_k}+F(\xi_i)=0,\quad
i=1,\ldots, n,
\label{6.3}
\end{eqnarray}
and the final expression for the eigenvalues $E_n(\lambda)$ is
given by the formula
\begin{eqnarray}
E_n(\lambda)=F^2(\lambda)+\frac{\partial}{\partial\lambda}F(\lambda)
+\sum_{i=1}^n \frac{F(\lambda)-F(\xi_i)}{\lambda-\xi_i}
\label{6.4}
\end{eqnarray}
Formulas (\ref{6.1}), (\ref{6.3}) and (\ref{6.4}) complete
the solution of problem (\ref{3.1}).
It is not difficult to see that this is only a partial
solution since the linear hull of vectors (\ref{6.1})
with fixed $n$ gives us only a certain negligibly small part of the
whole representation space defined in section \ref{2}. 
The above consideration enables one to conclude that the models which
we obtained are typical quasi-exactly solvable models.

\section{Simplest realizations of quasi-Gaudin algebra}
\label{7}

In previous sections we did not discuss the realizations of
algebra (\ref{2.1}). Now it is a time to do this.
In fact, we know only one realization which can be
constructed from the generators of standard Gaudin algebra
$\vec S(\lambda)$ satisfying the commutation relations (\ref{1.1}).
In order to construct it, we will need the special limiting operators
\begin{eqnarray}
S^\pm=\lim_{\lambda\rightarrow\infty}\lambda S^\pm(\lambda),\qquad
S^0=\lim_{\lambda\rightarrow\infty}\lambda S^0(\lambda),
\label{7.1}
\end{eqnarray}
which form the $sl(2)$ algebra. Using these operators, let us take
\begin{eqnarray}
S_n^-(\lambda)&=&S^-(\lambda)+\frac{n+f-S^0}{\lambda-a}\nonumber\\
S_n^0(\lambda)&=&S^0(\lambda)+\frac{n+f-S^0+b}{\lambda-a}\nonumber\\
S_n^+(\lambda)&=&S^+(\lambda)+\frac{n+f-S^0+2b}{\lambda-a}
\label{7.2}
\end{eqnarray}
where $a$ and $b$ are complex
parameters and $f=\lim_{\lambda\rightarrow\infty}\lambda
f(\lambda)$.
It can be easily checked by direct calculations that operators (\ref{7.2})
actually satisfy the quasi-commutation relations (\ref{2.1}).

The lowest weight functions $F(\lambda)$ and $G(\lambda)$ characterizing
the representation of this algebra read
\begin{eqnarray}
F(\lambda)=f(\lambda)+\frac{b}{\lambda-a},
\quad G(\lambda)=\frac{1}{\lambda-a}.
\label{7.3}
\end{eqnarray}

Now it is absolutely clear that algebra (\ref{2.1}) is a
deformation of the Gaudin algebra. The role of the
deformation parameter is played by number $a$. If
$a\rightarrow\infty$, then the $n$-dependence of the
operators (\ref{7.2}) disappears and they transform into
ordinary generators of Gaudin algebra. Respectively, the
quasi-commutators in (\ref{2.1}) become the ordinary ones. 
As to the formulas (\ref{7.3}) defining the representations
of algebra (\ref{3.1}), they, in the limit $a\rightarrow\infty$
also transform into the relations (\ref{1.4}) for Gaudin algebra.

\section{Quasi-exact solvability: a different point of view}
\label{8}

Substituting (\ref{7.2}) into (\ref{2.2}), we obtain the
form of the operators (\ref{2.2}):
\begin{eqnarray}
C_n(\lambda)&=&S^0(\lambda)S^0(\lambda)-\frac{1}{2}
S^-(\lambda)S^+(\lambda)-\frac{1}{2}S^+(\lambda)S^-(\lambda)+\nonumber\\
&+&\frac{2S^0(\lambda)(n+b+f-S^0)-S^-(\lambda)(n+2b+f-S^0)
-S^+(\lambda)(n+f-S^0)}{\lambda-a}+\nonumber\\
&+&\frac{b(b-1)}{(\lambda-a)^2}
\label{8.1}
\end{eqnarray}

Let us now check in an independent way that the operators (\ref{8.1})
actually describe quasi-exactly solvable models. First of
all, note that if the formulas (\ref{7.2}) hold, then the
representation spaces of Gaudin and our algebras coincide.
Denote by $\Phi_n$ the linear hull of all vectors 
$S^+(\xi_k)\ldots S^+(\xi_1)|0\rangle$ with arbitrary
$\xi_1,\ldots,\xi_k$ and $k\le n$. It is known that
if $f(\lambda)$ is a rational function, then 
$\dim\Phi_n<\infty$, for any $n$. From the obvious relations
$S^\pm(\lambda)\Phi_n\subset\Phi_{n\pm 1}$,
$S^0(\lambda)\Phi_n\subset \Phi_n$, $(n+f-S^0)\Phi_n \subset \Phi_{n-1}$
and $(n+f-S^0)\Phi_m\subset \Phi_m$ for $m\neq n$
it immediately follows that the operator $C_n(\lambda)$
admits only one algebraically constructable 
invariant subspace, $\Phi_n$. This subspace
is finite-dimensional and therefore the models (\ref{3.1})
are quasi-exactly solvable.

\section{Conclusion. From quasi-$G[sl(2)]$ to quasi-$sl(2)$}
\label{9}

It is known that the $sl(2)$ algebra can be considered as a
limiting case of the $sl(2)$-Gaudin algebra (\ref{1.1}). The generators
of the latter are given by the formula (\ref{7.1}).

It is natural to ask what kind of algebra will appear
if we consider an analogous limit of the deformed algebra 
(\ref{2.1})? In analogy with the Gaudin case, we can define the 
generators of this algebra by formulas
\begin{eqnarray}
S_n^\pm=\lim_{\lambda\rightarrow\infty}\lambda S_n^\pm(\lambda),\qquad
S_n^0=\lim_{\lambda\rightarrow\infty}\lambda S_n^0(\lambda).
\label{9.1}
\end{eqnarray}
Multiplying the quasi-commutation relations (\ref{2.1}) by $\lambda\mu$
and tending both $\lambda$ and $\mu$ to infinity we easily
derive the relations between these generators, which read
\begin{eqnarray}
S_n^0S_n^0
-S_n^0S_n^0=0, \quad
S_{n\pm 1}^\pm S_n^\pm
-S_{n\pm 1}^\pm S_n^\pm =0, \nonumber\\[0.3cm]
S_{n\pm 1}^0S_n^\pm
-S_n^\pm S_n^0=
\pm S_n^\pm, \quad
S_{n+1}^-S_n^+
-S_{n-1}^+S_n^-=2S_n^0.
\label{9.2}
\end{eqnarray}
We can consider (\ref{9.2}) as definig relations of
a certain modification of the $sl(2)$ algebra which can be called
``quasi-$sl(2)$ algebra''. 
Despite the fact that it {\it is not} a Lie algebra, it has
many properties similar to those of the ordinary sl(2).
For example, it has a quasi-analogue of the Casimir
operator,
\begin{eqnarray}
C_n=S_n^0S_n^0-\frac{1}{2}S_{n+1}^-S_n^+-
\frac{1}{2}S_{n-1}^+S_n^-
\label{9.3}
\end{eqnarray}
which quasi-commutes with all generators:
\begin{eqnarray}
C_nS_{n-1}^+=S_{n-1}^+C_{n-1},\quad
C_nS_{n+1}^-=S_{n+1}^-C_{n+1},\quad
C_nS_n^0=S_n^0C_n.
\label{9.4}
\end{eqnarray}
The representations of quasi-$sl(2)$ algebra can be
constructed in the same way as in quasi-Gaudin case.
Defining the lowest weight vector $|0\rangle$ by the formulas
\begin{eqnarray}
S_n^0|0\rangle=\left(F
+nG\right)|0\rangle,\qquad
S_n^-|0\rangle=nG|0\rangle,
\label{9.5}
\end{eqnarray}
where $F$ and $G$ are certain arbitrarily fixed numbers,
we can define the representation space $W_{F,G}$
as linear hull of vectors 
\begin{eqnarray}
|m\rangle_k=S_{m+k}^+S_{m-1+k}^+
\ldots S_{2+k}^+ S_{1+k}^+|0\rangle,
\label{9.6}
\end{eqnarray}
with arbitrary $k$ and $m$.  If we are interested in the
spectrum of quasi-Casimir operator, then it is easy to see
that, in contrast with the standard $sl(2)$ case, it is
not infinitely-degenerate and 
contains only a few number of exactly constructable eigenvectors.
For example, the only possibility for vector (\ref{9.6}) to be an 
eigenvector of (\ref{9.4}) is realized when $k=-1$ and $m=n$. In this
case, the corresponding eigenvalue is equal to $F(F-1)$.
So, one can say that the quasi-Casimir operators represent
the simplest (and, in some sense, trivial) quasi-exactly
solvable models.

It is remarkable that the generators of quasi-$sl(2)$
algebra can be realized in terms of the generators of ordinary $sl(2)$
algebra. The corresponding formulas can be obtained after
substitution of formulas (\ref{7.2}) into (\ref{7.1}) and read
\begin{eqnarray}
S_n^-=S^-+n+f-S^0,\quad
S_n^0=n+b+f,\quad
S_n^+=S^++n+2b+f-S^0.
\label{9.7}
\end{eqnarray}
This realization corresponds to the choice $F=f+b$ and $G=1$ in (\ref{9.5}).
We do not know at the moment if
there are other, more general, realizations of
quasi-$sl(2)$ algebra.

Concluding this paper, let us mention two problems which
arise as immediate consequences of the results obtained above. 
These are: 1) construction of general quasi-Gaudin algebras and models
and 2) construction of general quasi-Lie algebras.
However, one should stress again that our main goal 
is to develop a theory of general quasi-YB algebras. 
The first step in this direction will be
done in our next publication which will appear soon in
hep-th archive and in which we intend to
construct quasi-exactly solvable deformations of the 
inhomogeneous XXX spin chain and investigate the associated quasi-Yangian
${\cal Y}[sl(2)]$.

\end{document}